\documentclass[twocolumn,tighten]{aastex631}

\usepackage{graphicx}
\usepackage[utf8]{inputenc}
\usepackage{amsmath,amssymb}
\usepackage{epstopdf}
\usepackage{hyperref}

\newcommand{\Fref}[1]{Fig.~\ref{#1}}
\newcommand{\Eqref}[1]{Eq.~(\ref{#1})}
\newcommand{\be}{\begin{equation}}
\newcommand{\ee}{\end{equation}}
\newcommand{\bal}{\begin{align}}
\newcommand{\eal}{\end{align}}
\newcommand{\bear}{\begin{eqnarray}}
\newcommand{\eear}{\end{eqnarray}}
\newcommand{\nn}{\nonumber}

\newcommand{\md}{\mathrm{d}}

\newcommand{\kb}{k_{_B}}

\newcommand{\Obs}{Van_Doorsselaere:11,Wang:15,Prasad:18,Vashalomidze:19}
\newcommand{\Old}{Totten:95,Basu:04}

\begin{document}

\title{Influence of Ionization on the Polytropic Index of the Solar Atmosphere within Local Thermodynamic Equilibrium Approximation}

\author[0000-0001-5394-3027]{Albert~M.~Varonov}
\affiliation{Georgi Nadjakov Institute of Solid State Physics, Bulgarian Academy of Sciences,\\
72 Tzarigradsko Chaussee Blvd., BG-1784 Sofia, Bulgaria}

\author[0000-0002-6945-9632]{Todor~M.~Mishonov}
\affiliation{Georgi Nadjakov Institute of Solid State Physics, Bulgarian Academy of Sciences,\\
72 Tzarigradsko Chaussee Blvd., BG-1784 Sofia, Bulgaria}

\correspondingauthor{Todor~M.~Mishonov}
\email{mishonov@gmail.com}

\date{22 Dec 2023, winter solstice}


\begin{abstract}
An initial theoretical attempt to explain the observed decrease of the polytropic/adiabatic index $\gamma$ in the solar corona has been accomplished.
The chemical reactions of the ionization-recombination processes in local thermodynamic equilibrium (LTE) of a solar plasma cocktail containing heavy elements are found to cause $1.1 < \gamma \leq 5/3$ in the quiet solar atmosphere.
It is also shown that the quiet solar atmosphere is in LTE justifying this theoretical study.
This result is obtained by numerically solving the Saha equation and subsequently using a newly derived equation for calculation of the polytropic index from thermodynamic partial derivatives of the enthalpy and pressure with respect to density and temperature.
In addition, a comparison between calculated in this way polytropic index and measured from spectroscopic observations of propagating slow magneto-hydrodynamics (MHD) waves in coronal loops shows that LTE ionization  accounts for very small part of the observed decrease of $\gamma$ meaning that the solar plasma in the active region is not in LTE as expected.
However, the observed dependency of higher polytropic index at higher temperatures is confirmed by the current theoretical approach.
It is concluded that to account for the polytropic index decrease in the active regions of the solar corona, it is necessary kinetic non-LTE ionization calculations to be performed.
\end{abstract}

\section{Introduction}

Recent observations of magneto-hydrodynamic (MHD) waves propagating in the solar corona have been used to determine the polytropic (or adiabatic) index $\gamma$ of the solar coronal plasma.
The obtained values from several measurements vary between 
1 and  the mono-atomic value of the polytropic index $\gamma_a \equiv 5/3$~\citep{\Obs}.
Previous observations of the solar wind far from the Sun also show similar deviations of $\gamma$ from its mono-atomic value~\citep{\Old}.
Moreover, data from observations of magnetic field fluctuations in umbral flashes leads to measurement of the polytropic index via the magnetic pressure~\citep{Houston:18} which makes at least 2 recently devised methods for the determination of $\gamma$.

The knowledge of its value is of high physical interest and it's been researched for over 50 years~\citep{Petrie:07}.
This research actively continues as observed data of the solar wind from Parker Solar Probe~\citep{Nicolaou:23}, Ulyssess~\citep{Nicolaou:20}, Wind~\citep{Dayeh:22}, ACE~\citep{Poedts:11}, Cluster~\citep{Pang:22} spacecrafts are used to determine the polytropic index.
Knowing the nature of the deviations of the polytropic index and the physical processes causing it is a significant step forward in the study of the thermodynamics and energy balance of not only the solar and stellar atmospheric plasma, but also for cold proto-stellar molecular clouds~\citep{Donkov:21}.
This step allows better understanding of the numerous physical phenomena occurring in astrophysical plasmas,
MHD wave propagation and damping, heating and cooling processes both in quiet, eruptive or turbulent cases~\citep{Livadiotis:23}.

The main goal of the presented study in this work is to theoretically explain the measured deviation of $\gamma$ for the solar plasma and provide a reliable means for its numerical calculation from known mass density $\rho$ and temperature $T \equiv \kb T^\prime$,
where $T$ is energy units, $\kb$ is the Boltzmann constant and $T^\prime$
is the temperature in degrees.
Similar deviation of $\gamma$ has been recently found to be caused by the ionization-recombination processes by \citet{PhysA,ApJ:21}, where
an analytical formula in terms of the ionization degree for pure hydrogen (H) plasma has already been derived from the Saha ionization equation~\citep{Saha:21,LL5} in local thermodynamic equilibrium (LTE).
Here follows the continuation of this study where helium (He) and heavy elements have been included in the Saha equation.
Equations for numerical calculation of the ionization degree and the polytropic index $\gamma$ are derived and used for comparison with the available measurements.

Summarizing all of the above, the described in this paper approach studies the variations of the adiabatic polytropic index only due to the simplest chemical reactions ionization and recombination via the Saha equation of the 10 most abundant chemical elements in the Solar atmosphere.
These ionization-recombination processes in LTE lead to variations of the enthalpy and pressure resulting in deviations of $\gamma$ from 5/3.
Mass density $\rho$ and temperature $T$ are assumed to be known and are starting parameters for the calculations while heating is taken to be zero.

Similar study has been performed by \citet{Baturin:22} for the solar interior again solving the Saha equation.
An excellent comparison between both methods for the solution of the Saha equation shows that the described here results can also be used within the interior of the Sun.
However, the focus of this study remains the solar atmosphere and the available observed and measured parameters there.

\section{The Saha Equation for Chemical Equilibrium}

Following \citet[Eqs.~(101.1)--(101.4)]{LL5}, the Saha equation can be written as
\begin{align}
& r_{i,a}  =\frac{g_{i, a} \, g_e}{g_{{i-1}, a}}
\frac{n_{s,i,a}}{n_e} r_{i-1,a},
\label{Saha} \\
&
n_{s,i,a}\equiv n_q\exp(-I_{i,a}/T),\qquad
n_q\equiv \left(\frac{mT}{2\pi\hbar^2}\right)^{\! 3/2},\nn
\end{align}
where $r_{i,a}$ are relative concentrations of each ion $n$ of each chemical element $a$ in the cocktail,
index $i=0, \dots, Z_a$ denotes ionization level and
index $a$ denotes chemical element.
The ionization energy is $I_{i,a}$
where $I_{0,a}=0$ since $i=0$ denotes the atoms (0-th ionization level),
$g_{i,a}$ is the statistical weight of the atoms and ions,
$g_e$ is the electron statistical weight,
$n_e$ is the electron density,
$m$ is the electron mass and $\hbar$ is the Planck constant.
The relative concentrations for each chemical element are all normalized
\be
r_a=\sum_{i=0}^{Z_a}r_{i,a},
\quad 
r_{i,a}=\frac{r_{i,a}}{r_a},
\quad 
i=0,\,\dots,\, Z_a. \nn
\ee
so that
\be
\sum_{i=0}^{Z_a}r_{i,a}=1. \nn
\ee
The following additional notions need to be introduced
\begin{align}
&
n_e=n_\rho\, \mathcal N_e,\quad 
\mathcal N_e=\sum_{a} \overline a_a \nu_a, \quad 
\nu_a=\sum_{i=1}^{Z_a} i \, r_{i,\,a},
\label{Electron_Density} \\
&
\mathcal{N}_\mathrm{atom}=\sum_{a}\overline{a}_a,
\qquad
\mathcal{N}_e=\sum_{a}\overline{a}_a\sum_{i=1}^{Z_a} i \, r_{i,a}.\\
&
\mathcal{N}_\mathrm{tot}=
\mathcal{N}_\mathrm{atom}+\mathcal{N}_e,
\quad n_\mathrm{tot}=\left(\mathcal{N}_e+\mathcal{N}_\mathrm{atom}\right)n_\rho,\\
&
n_\rho = \rho/M^*, \quad M^*=\sum_a \overline a_a M_a,\;\;
\langle M\rangle\equiv M^*/\mathcal{N}_\mathrm{tot},
\end{align}
where $\overline{a}_a$ is the abundance of each chemical element,
$n_\rho$ is the density of all atoms and ions or all particles with mass,
here the electron mass is neglected since $m/M \approx 1/1836$,
$M$ is the proton mass
and $n_\mathrm{tot}$ is the total density of all particles.
$\mathcal{N}_\mathrm{tot}$, $\mathcal{N}_\mathrm{atom}$, $\mathcal{N}_e$
are respectively the number of total, atoms and ions, electrons per unit density,
while $\nu_a$ is the number of electrons the chemical element $a$ has given to the cocktail per unit density.
One can think of labeling the electrons and then $\nu_a$ is the number of electrons per unit density with label $a$.

\section{Equation for Calculation of the Polytropic Index}

The calculated densities of the plasma atoms, ions and electrons allows the calculation of the enthalpy per unit mass
\begin{align}
&
w=\frac1{M^*}
\sum_{a}\overline a_a\left[c_p(1+\nu_a)T+\overline \epsilon_{a}\right],
\label{w} \\
&
\overline \epsilon_{a}\equiv\sum_{i=1}^{Z_a} i \, r_{i,a} J_{i,a},
\qquad
J_{i,a}\equiv\sum_{m=1}^i I_{m,a} \nn
\end{align}
and pressure
\be
p=n_\mathrm{tot}T=n_\mathrm{\rho}\mathcal{N}_\mathrm{tot}T.
\label{p}
\ee
For the internal energy per unit mass $\varepsilon$ we have to substitute $c_p$ with $c_v$.
The partial derivatives of $w$ and $p$ with respect to $\rho$ and $T$ participate in a newly derived formula for the calculation of the polytropic index via the Jacobian
\begin{align}
\mathcal{J} \equiv &
\frac{\partial (w,p)}{\partial (T,\rho)} 
=
\left(\frac{\partial w}{\partial T}\right)_{\!\! \rho}
\left(\frac{\partial p}{\partial \rho}\right)_{\!\! T}
-\left(\frac{\partial w}{\partial \rho}\right)_{\!\! T}
\left(\frac{\partial p}{\partial T}\right)_{\!\! \rho}, 
\label{Jacobian} \\
\mathcal{C}_v\equiv &
\left(\dfrac{\partial w}{\partial T}\right)_{\!\! \rho}
-\dfrac{1}{\rho}\left(\dfrac{\partial p}{\partial T}\right)_{\! \! \rho},
\qquad \gamma
=\frac{\rho}{p}\cdot
\dfrac{\mathcal{J} }{\mathcal{C}_v}.
\label{gamma}
\end{align}
The derivation of the heat capacity per unit mass is given in 
the Appendix \Eqref{partial_W}.
For the limiting case of an ideal gas without of chemical reactions
meaning $r_{i,a}=\mathrm{const}$
\begin{align}&
p=n_\rho \mathcal{N}_\mathrm{tot}T, \quad
w=\frac{c_pT}{\langle M\rangle}  +\mathrm{const}, \quad
\rho=n_\rho M^*
\nn\\
&
\varepsilon=\frac{c_vT}{\langle M\rangle}  +\mathrm{const}, \quad
\mathcal{C}_v=\left(\frac{\partial \varepsilon}{\partial T}\right)_{\! \!\rho}
=\frac{c_v}{\langle M\rangle},
\nn
\end{align}
the Jacobian~\Eqref{Jacobian} 
\be
\mathcal{J} = c_p \frac{T}{\langle M\rangle^2}
\nn
\ee
and \Eqref{gamma} gives the mono-atomic value for the polytropic index
\be
\gamma = c_p/c_v = 5/3 = \gamma_a
\label{boundary}
\ee
where the relation denoting  the Newtonian sound speed $c_\mathrm{_N}$
\be
\frac{p}{\rho}=\frac{n_\mathrm{tot}}{M^*n_\rho}=\frac{T}{M^*/\mathcal{N}_\mathrm{tot}}
=\frac{T}{\langle M\rangle}=c_\mathrm{_N}^2
\ee
has been used.
The derivation of the equation for the polytropic index \Eqref{gamma} is given in Appendix~\ref{sec-gamma}.

Few notes deserve to be included here.
The equations for enthalpy \Eqref{w} and pressure \Eqref{p} have only terms related to the number of particles in the cocktail.
And the partial derivatives of $w$ and $p$ with respect to $T$ and $\rho$ determine the value of $\gamma$ in \Eqref{gamma}, meaning that the ionization-recombination processes change the enthalpy and the pressure which in turn lead to variations of the adiabatic polytropic index with no heating.

\subsection{Details about the Numerical Calculations}

The Saha equation \Eqref{Saha} is a transcendental one since for calculation of the relative concentrations $r_{i,a}$ it is necessary to know the electron density $n_e$,
while at the same time these concentrations are used to calculate $n_e$ from \Eqref{Electron_Density}.
As a mathematical problem the Saha equation gives a system of non-linear equations that need to be solved with high accuracy.
All these variables are described by a single parameter -- the electron density $n_e$.
The whole iterative procedure is distinctively shortened by using of a reliable
numerical method for extrapolation towards infinite amount of iterations based on the Wynn identity~\citep{Wynn:66}.
This identity at the same time also gives an empirical evaluation of the calculation error which is comparable with the machine epsilon~\citep{Pade}.
The successive iterations start with $\rho$ and $T$ as input parameters, while
the initial electron density is chosen $n_e^0 \sim n_\rho$.
After calculation of the new electron density $n_e$ with the Saha equation,
$n_e^0 = n_e$ (new becomes old)
and again a new calculation is performed.
This procedure is repeated several times and with all obtained values for $n_e$ a Wynn-epsilon algorithm for acceleration convergence~\citep{Pade} is used to calculate the final value.
The algorithm has a reliable error calculation which is used for a decision whether the obtained value of $n_e$ to be accepted or a new series of calculation to be performed. 
All this allows the calculation of the Saha equation to be performed on an ordinary personal computer using FORTRAN.
This of course is for the present case of the ionization energies of the chemical elements in the plasma cocktail only taken from~\citet{NIST_ASD}.
For the inclusion of the electron transition energies a much larger computing hardware would be certainly necessary.

\section{Results and Discussion}

The main goal of this study is to explain obtained data of electron density $n_e$ and polytropic index $\gamma$ in the inner Solar atmosphere.
This data is a result of observations of both quiet solar atmosphere and active loop regions.
The chemical elements included in this calculation are 10:
H, He, O, C, N, Ne, Mg, Si, Fe, S with their abundances $\overline{a}_a$ taken from \citep[Table~2]{Avrett:08} and
as mentioned earlier, their ionization levels from~\citet{NIST_ASD}.
A starting value of $r_{0,a} = 10^{-64}$ is chosen due to $n_q$ becoming extremely large for higher temperatures.

\subsection{Quiet Solar Atmosphere}

\begin{figure}[ht]
\centering
\includegraphics[scale=0.42]{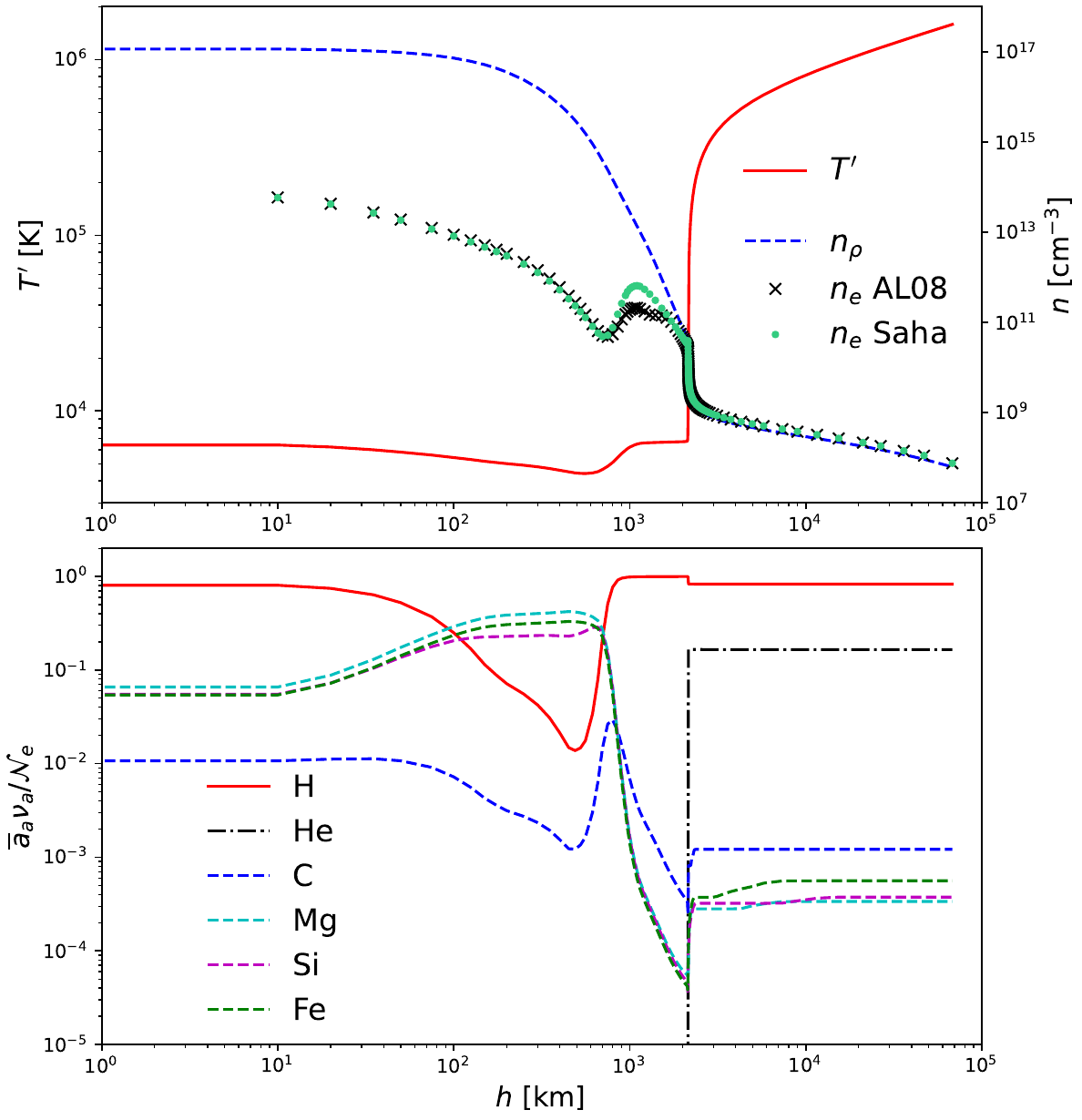}
\caption{Top: height profile of the quiet solar atmosphere by \citet[Model C7, Table~26]{Avrett:08} with $T^\prime$, total hydrogen density $n_\rho$, $n_e$ AL08 ($\times$),
together with $n_e$ calculated from \Eqref{Saha} Saha (\textcolor{green}{$\bullet$}).
Below: the relative contributions of some of the chemical elements in the cocktail to $n_e$ divided by $\mathcal{N}_e$.
The agreement between the semi-empirical non-LTE model and the theoretical LTE Saha calculation is remarkable and its correlation is shown in \Fref{corr-AL08}.
The only visible deviation between semi-empirical and theoretical values is at the maximum C and S (not shown here for brevity) contribution to $n_e$, pointing that most probably more lines from these elements need to be included.
}
\label{AL08}
\end{figure}

The basis of the results in the quiet solar atmosphere is the semi-empirical model quiet-Sun model by \citet[Model C7, Table~26]{Avrett:08} (AL08).
This model is a result optically thick non-LTE radiative transfer calculations  H, C, O and other chemical elements lines and continua of taken from the SUMER atlas \citep{SUMER} and the atlas from HRTS by \citet{HRTS}.
It is a height dependent one-dimensional profile shown in \Fref{AL08} top
with $h=0$ denoting the solar photosphere and where the following most relevant parameters are shown:
$T^\prime$ is temperature in Kelvins,
$n_\rho$ is total hydrogen density, which is different from the already introduced mass particles density and
$n_e$ the density of electrons with $\times$ and named AL08.
Alongside the profile parameters, the numerical calculation of $n_e$ with the Saha equation \Eqref{Saha} using $n_\rho$ and $T^\prime$ from AL08 is also shown with \textcolor{green}{$\bullet$}.
Below the profile, the relative contributions of some of the chemical elements in the cocktail to $n_e$ is shown.
These contributions are normalized so that $\mathcal{N}_e=1$ to alleviate the current analysis.
The juxtaposition of both top and below graphs helps to determine where the largest deviation between $n_e$ from AL08 and the Saha equation is.
It occurs at approximately the same position as the C and S (not shown here) ionization contribution maximums just below the solar transition region (TR).
This is a clue that some C and S transition lines need to be included for better correlation.
The contribution of all chemical elements to the ionization is described in App.~\ref{sec-ion}.

Another interesting discovery is the domination of the Mg, Fe and Si in the ionization of the upper solar chromosphere.
Actually this is to be expected since these elements way more easily than H give 1-2 electrons and in the local temperature minimum where H is so weakly ionized that these 3 much less abundant chemical elements determine the ionization of this region of the solar chromosphere.

\begin{figure}[ht]
\centering
\includegraphics[scale=0.5]{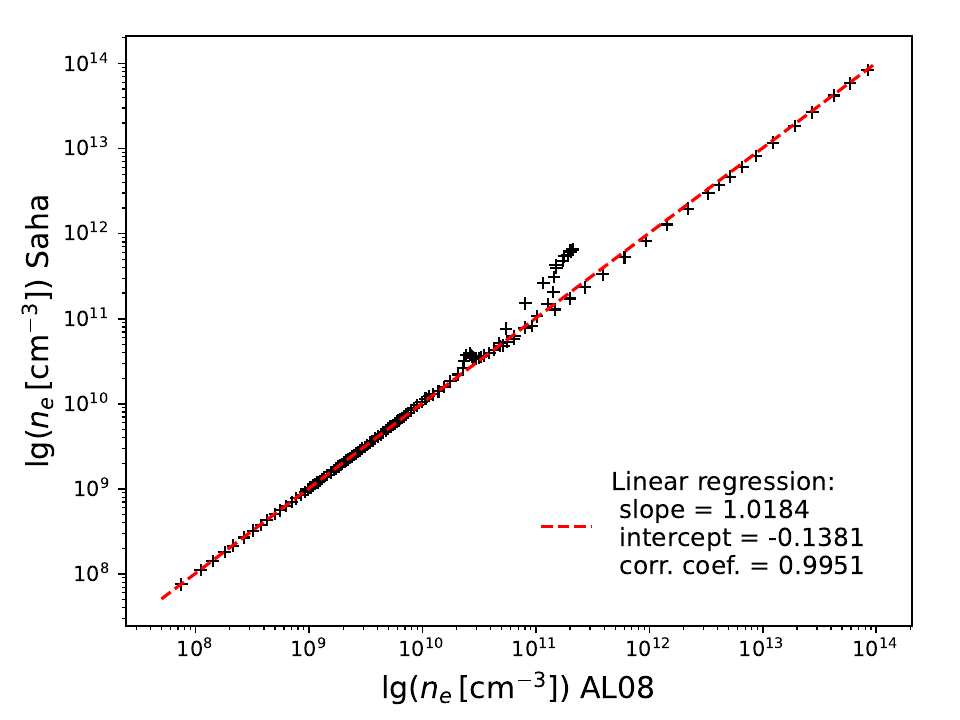}
\caption{Linear correlation between $n_e$ from AL08 ($x$ axis) and $n_e$ from the Saha equation \Eqref{Saha} ($y$ axis) shown in \Fref{AL08} (top).
The slope of the line = 1.0184 with a correlation coefficient = 0.995 meaning that the difference between the semi-empirical and theoretically calculated electron density is $< 2$\% with almost $3 \sigma$ confidence level.
The significant correlation between observed the electron density
and the calculated by Saha equation reveals the local 
thermodynamic equilibrium (LTE) is an acceptable initial approximation.
} 
\label{corr-AL08}
\end{figure}

The log-log graph of both AL08 and Saha in $n_e$ \Fref{AL08} (top) shows excellent correlation between them and its linear regression of this correlation is shown in \Fref{corr-AL08}.
The deviation between both electron density profiles in \Fref{AL08} is also clearly seen here in the dispersing points from the regression line.
This deviation leads to less to a slope of 1.0184, less than 2\% larger than 1 with 0.995 correlation coefficient.
One can hardly ask for a better correlation in astrophysics where the remote observations and their subsequent processing are difficult and consume lots of time and resources.
A scientist's access to the space lab is more remote than could ever be,
the true definition of work from home.

\begin{figure}[ht]
\centering
\includegraphics[scale=0.5]{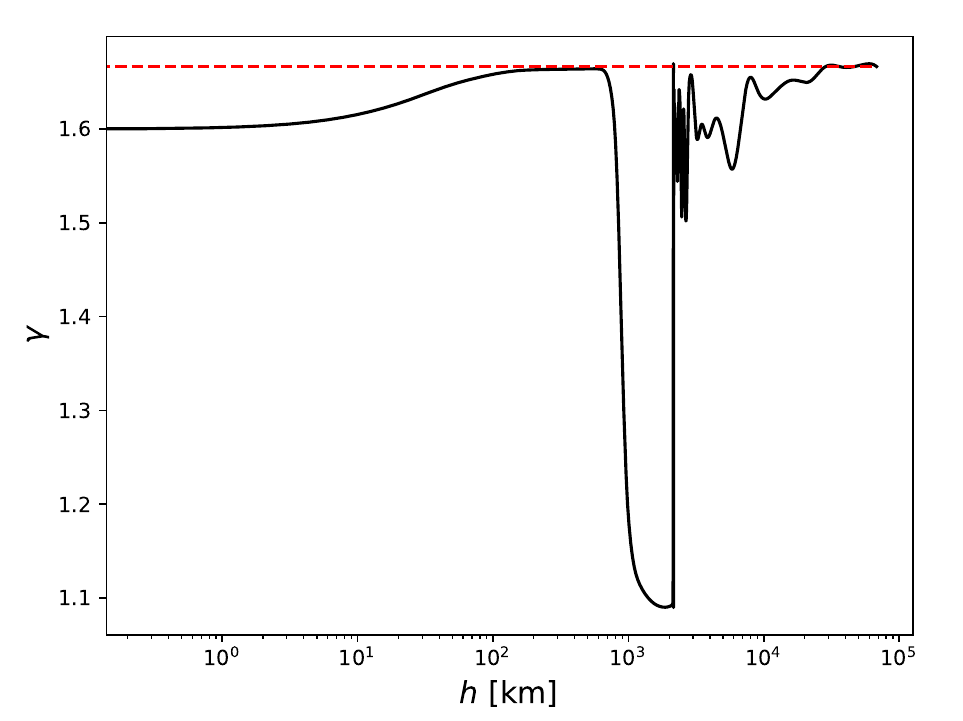}
\caption{The polytropic (adiabatic) index $\gamma$ in the quiet solar atmosphere according to the \citet{Avrett:08} profile (solid line),
dashed line denotes $\gamma_a=5/3$.
The values for which $\gamma \gtrapprox \gamma_a$ are artificial effects from the splicing of the curve, no calculated value exceeds 5/3.
Comparing with \Fref{AL08} the large decrease of $\gamma$ is due to the H ionization and the oscillations right after it are due to the step-like ionization of He.
The deep minimum in the upper chromosphere is related to the hydrogen ionization at almost neutral helium.
} 
\label{gamma-AL08}
\end{figure}

After reliably obtaining $n_e$, the enthalpy $w$ \Eqref{w}, the pressure $p$ \Eqref{p} and their partial derivatives are calculated.
Having all these, the Jacobian $\mathcal{J}$ \Eqref{Jacobian} and finally the polytropic index $\gamma$ \Eqref{gamma} are calculated.
The polytropic index for the AL08 profile is shown in \Fref{gamma-AL08}.
The largest $\gamma$ decrease down to $\approx 1.1$ occurs approximately at the distance where the H ionization starts increasing, see \Fref{Saha}.
The following oscillations are correlated with the subsequent step-like increase of the He ionization.
At low heights $\gamma \approx 1.6$ and starts increasing with H ionization decreasing and in the dominant ionization region of Mg, Fe and Si  $\gamma \approx \gamma_a$ meaning that the level of ionization of this region is very low.
Farthest from the Sun again $\gamma \approx \gamma_a$ but this time it is because of the almost full ionization of all the elements, only Fe has succeeded in keeping its last 2$s$ electrons (FeXXV).
This full ionization and Fe partial cause the gradual oscillations tending to 5/3.
This boundary value is in full agreement for the boundary case without chemical reactions \Eqref{boundary}, the last 2 Fe ionization levels are $\approx 4 \times$ larger than the one preceding them.

Unfortunately, the authors could not find any calculations based on spectroscopic measurements of the polytropic index for the quiet Sun.
Hope this work to inspire astrophysicists to perform such measurements or at least obtain estimates for the quiet solar atmosphere.
It seems for now the interesting phenomena are within the active regions and comparison with measurements of $\gamma$ in loops follows next.

\subsection{Active Loop Regions}

The only available obtained data for $\gamma$ in the solar atmosphere that could be found are from spectroscopic measurements of coronal loops.
First such observation by \citet{Van_Doorsselaere:11}
measured equilibrium $T^\prime=1.7$~MK and $n_e=1.7 \times 10^{15}~\mathrm{m}^{-3}$ and obtaining $\gamma = 1.10 \pm 0.02$.
Followed by \citet[Figs. 4 and 5]{Wang:15} with 1 measurement 
 $T^\prime=8.7$~MK and $n_e=2.6 \times 10^{9}~\mathrm{cm}^{-3}$ giving $\gamma = 1.64 \pm 0.08$ and after removal of the slowly varying trend
 $T^\prime=12.6$~MK and $n_e=2.7 \times 10^{9}~\mathrm{cm}^{-3}$ giving $\gamma = 1.66 \pm 0.09$.
 A systematic study of several loops and more than 30 obtained values for the polytropic index was made by \citet[Table~1]{Prasad:18} based on measurements of $T^\prime$ and $n_\rho$.
And finally \citet{Vashalomidze:19} during solar coronal rain measured $T^\prime \approx 2-4$~MK and $n_e \approx 1.35 \times 10^{9}~\mathrm{cm}^{-3}$ and obtaining $\gamma = 1.30 \pm 0.06$.
At the onset of the rain $\gamma = 2.11 \pm 0.11$ which falls outside the scope of this study.

\begin{figure}[ht]
\centering
\includegraphics[scale=0.5]{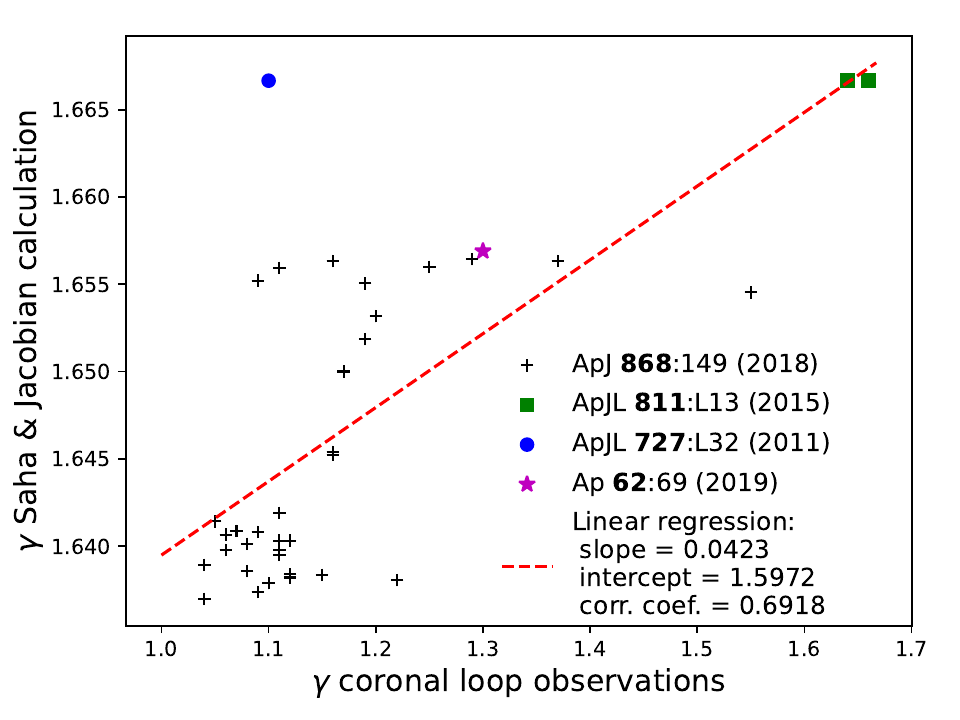}
\caption{Correlation of the polytropic index $\gamma$ obtained from coronal loops spectroscopic observations and calculated from the Saha equation and the Jacobian.
The abbreviations of the papers denote:
ApJL \textbf{727}:L32 (2011) for \citet{Van_Doorsselaere:11},
ApJL \textbf{811}:L13 (2015) for \citet{Wang:15},
ApJ \textbf{868}:149 (2018) for \citet{Prasad:18},
Ap  \textbf{62}:69 (2019) for \citet{Vashalomidze:19}.
The correlation coefficient of 0.69 means good agreement, 
however the slope of 0.043 points that non-LTE processes are involved in the adiabatic index decrease in the active regions.} 
\label{gamma-Obs}
\end{figure}

Having $T$ and $n_\rho$ or $n_e$ ($n_\rho \approx n_e$ for starter)
the adiabatic index $\gamma$ can be easily calculated with the Saha equation \Eqref{Saha} and via the Jacobian \Eqref{gamma}.
The comparison of the obtained from spectroscopic measurement results and the calculated from LTE are depicted in \Fref{gamma-Obs}.
The observed $\gamma$ varies between 1.0 and $\gamma_a$, while the calculated $\gamma$ between 1.664 and $\gamma_a$, the LTE theoretical calculation accounts for very small decrease of the polytropic index from the mono-atomic one. 
This indicates that most probably non-LTE processes cause the substantial decrease of the adiabatic index in the active regions.
The high correlation coefficient of the linear regression of 0.69 shows the relation between the ionization and these non-LTE processes.
This is a promising clue the observed $\gamma$ decrease might be caused by sudden ionization of the chemical elements due to high mass influxes and temperature spikes.
In other words, a kinetic approach of eruptive events in the solar atmosphere should be the next step towards the explanation of the observed polytropic index decrease in the coronal loops.

Finally it is worth noting the only observation by \citet{Wang:15} denoted by ApJL \textbf{811}:L13 (2015) in \Fref{gamma-Obs} matching with its corresponding theoretically calculated one $\gamma = 1.666\dots$.
For such a hot solar plasma it is possible full ionization to be in place, i.e. all electrons are free and this is exactly the case of an ideal gas with no chemical reactions occurring, at these temperatures the recombination processes are highly unlikely.
This strengthens the above stated proposition for a kinetic study of non-LTE ionization processes in solar eruptive events and active regions.

The data behind the figures presented in this study can be found at \dataset[10.5281/zenodo.10401574]{\doi{10.5281/zenodo.10401574}}.

\section{Conclusions}

This theoretical study unambiguously shows that the quiet solar atmosphere is in LTE.
The very high agreement between the electron density $n_e$ obtained from the semi-empirical non-LTE model by \citet{Avrett:08} and from the Saha equation calculation is quite astonishing in the field of astrophysics.
The tremendous systematic research of the quiet Sun from \citet{VAL:81} to \cite{Avrett:08} has resulted in LTE solar atmospheric height profiles without implying any equilibrium conditions.
The Saha equation \Eqref{Saha} successfully repeats this profile meaning  it can be safely used in both MHD and kinetic numerical calculations of quiet solar and stellar atmospheres where heavy elements Z should be included.
The described here solution can be easily incorporated in any numerical source code and takes several machine seconds time for a contemporary personal computer without any parallelization.

Alongside the ionization-recombination in LTE, a new equation for the calculation of the polytropic index \Eqref{gamma} via Jacobian \Eqref{Jacobian} of thermodynamic partial derivatives of the entropy $w$ and pressure $p$ with respect to density $\rho$ and temperature $T$ has been derived.
Together with the Saha equation, it allows numerically to calculate the polytropic index $\gamma$ of an arbitrary chemical cocktail where equilibrium ionization-recombination processes occur.
Applied to the quiet Sun height profile of \citet{Avrett:08} it is shown that
$1.1 < \gamma \leq 5/3$ \Fref{gamma-AL08}.
This is a new result needing observational verification at least after the solar TR.

Observations of slow MHD propagating waves in coronal loops of solar active regions by \citet{\Obs} confidently show that
$1.04 \leq \gamma \leq 1.66$ with a single exception from \citet{Vashalomidze:19}.
Using the determined from the observations $n_\rho$ and $T$,
$\gamma$ for each observational measurement has been theoretically calculated.
The comparison between measured from observation and theoretically calculated polytropic indices shows that the LTE calculation accounts for very small decrease of $\gamma$ \Fref{gamma-Obs}.
On other hand, there is significant correlation meaning that indeed  ionization could be causing this decrease, albeit in non-LTE.
The only match between observed and calculated polytropic index occurs at very high temperature close to 10~MK.
A plausible explanation of this match is the almost full ionization of the whole chemical cocktail leading to $\gamma \approx 5/3$.
The conclusion of \citet{Prasad:18} that higher $\gamma$ corresponds to higher temperatures is fully consistent with our results.
Despite non-accounting for the decrease of $\gamma$ in active regions, our study reveals what might probably be the cause of this:
sudden ionization processes.
Their non-LTE kinetic treatment which is beyond the scope of this study should be put next on the agenda of the polytropic index in the solar corona problem.

To summarize, at fixed temperature the electron density determines the equilibrium concentrations of the different ion states of the elements.
The high correlation between the calculated and the observed electron densities show that within acceptable accuracy the Saha equation describes the ion height distribution.
And in this way regardless of the heating mechanism, LTE is an acceptable initial approximation for further analysis of the processes in the solar atmosphere.

The described approach for determination of the variations of the adiabatic polytropic index as a result from the chemical ionization-recombination processes via the Saha equation is applicable to various astrophysical, physical and chemical processes taking place in LTE or near-LTE.
Such processes take place in stellar atmospheres of slowly rotating low activity stars, molecular clouds, star forming regions, for instance.
As more (or less) chemical elements can be used in a calculation, so additional energy levels, activation, mixing, vaporization, chemical bond, molecular vibrational and rotational energies can be added, depending on the composition, mass density and temperature of the chemical cocktail.
However, it should be noted that with increase of the chemical elements and their respective energy levels, the calculation quickly becomes quite complex and time consuming even for a high-performance computers and therefore we advise that a reasonable evaluation and analysis of the dominant elements and energy levels to be made beforehand.
Similar analysis has been done in Sec.~\ref{sec-ion} illustrating each chemical element contribution allowing to further fine tune the calculation if necessary.

\section*{Acknowledgments}

\begin{acknowledgments}
The authors are grateful to
Claude~Brezinski for the discussion related to the practical realization of the Wynn algorithm,
Yana Maneva for the discussion of the applicability of the Saha equation for the solar atmosphere,
Theodora~Nikolova, Atanas~Batinkov and Emil~Petkov for their interest in the research.
The authors are also thankful to Hassan Chamati for the hospitality in ISSP, BAS
where this study was performed.
This work is supported by Grant KP-06-N58/1 from 15.11.2021 of the Bulgarian National Science Fund.
\end{acknowledgments}

\clearpage

\appendix

\section{Derivation of the equation for the polytropic index}
\label{sec-gamma}

This section is devoted to the derivation of \Eqref{gamma}.
Since the focus of the manuscript is the theoretical approach for explanation of measurements obtained from astrophysical observations, the thermodynamic derivation comes second in this appendix section.
If necessary, it can easily be moved to a main section or completely removed.

Originally the authors derived \Eqref{gamma} by working on a numerical method for solution of the MHD equations for the density $\rho$ and temperature $T$ height profiles of the solar chromosphere but
here a ``more physical'' thermodynamic proof is presented.
A transition from hydrodynamic to thermodynamic notions for constant mass of the fluid particle 
\be
w=\frac{W}{M},\qquad
s=\frac{S}{M},\qquad
\mathcal C_v=\frac{C_v}{M},\qquad
p=P,\qquad
\rho=\frac{M}{V},\qquad
\md \rho=-\frac{M}{V^2}\md V.
\ee
leads that Jacobian \Eqref{Jacobian} can be represented as 
\be
\mathcal{J} =
\frac{\partial (w,p)}{\partial (T,\rho)} 
=\frac1{\rho^2}
\frac{\partial (W,P)}{\partial (T,V)} 
=
\frac{1}{\rho^2} 
\left [ 
\left(\frac{\partial W}{\partial V}\right)_{\!\! T}
\left(\frac{\partial P}{\partial T}\right)_{\!\! V}
-
\left(\frac{\partial W}{\partial T}\right)_{\!\! V}
\left(\frac{\partial P}{\partial V}\right)_{\!\! T}
\right ].
\ee
Substituting here the thermodynamic relation 
or the enthalpy $W$ partial derivatives~\citep[Eqs.~(16.7)  and (16.8)]{LL5}
\be
\left(\frac{\partial W}{\partial V}\right)_{\!\! T}
=T\left(\frac{\partial P}{\partial T}\right)_{\!\! V}
+V\left(\frac{\partial P}{\partial V}\right)_{\!\! T},
\qquad
\left(\frac{\partial W}{\partial T}\right)_{\!\! V}=C_v+V\left(\frac{\partial P}{\partial T}\right)_{\!\! V}
\label{partial_W}
\ee
we obtain
\be
\mathcal{J} = \frac{1}{\rho^2}  
\left [ 
 T\left(\frac{\partial P}{\partial T}\right)^{\!2}_{\!\! V}
-C_v \left(\frac{\partial P}{\partial V}\right)_{\!\! T} 
\right ].
\ee
Another substitution  for the partial derivative of the pressure $P$ with respect to $V$ for constant $T$~ \citep[Eq.~(16.16)]{LL5}
\be
\left(\frac{\partial P}{\partial V}\right)_{\!\! T} 
= \left ( \frac{\partial P}{\partial V} \right )_{\!\!S} 
+ \frac{T}{C_v} \left( \frac{\partial P}{\partial T} \right)^{\!2}_{\!\!V} 
\ee
leads to
\be
\mathcal{J} = -\frac{C_v}{\rho^2} \left ( \frac{\partial P}{\partial V} \right )_{\!\!S}.
\label{Jacobian_TD}
\ee
The denominator of \Eqref{gamma} in thermodynamic notions is
\be
\left(\dfrac{\partial w}{\partial T}\right)_{\! \rho}
-\dfrac{1}{\rho}\left(\dfrac{\partial p}{\partial T}\right)_{\! \! \rho} =
\frac{1}{M} 
\left [
\left(\dfrac{\partial W}{\partial T}\right)_{\! \!V} - V \left(\dfrac{\partial P}{\partial T}\right)_{\! \! V}
\right ] = \frac{C_v}{M},
\label{denom_TD}
\ee
where the left equation in \Eqref{partial_W} has been used.
A substitution of the Jacobian \Eqref{Jacobian_TD} and \Eqref{denom_TD} in thermodynamic notions in \Eqref{gamma}
yields
\be
\gamma =\frac{\rho}{p} \left [ -\frac{M}{\rho^2}   \left ( \frac{\partial P}{\partial V} \right )_{\!\!S} \right ] 
= \frac{\rho}{p}  \left ( \frac{\partial p}{\partial \rho} \right )_{\!\!s} 
\ee
which can be arranged in the much more familiar form of the equation for sound speed $c_0$ of an ideal gas
\be
 \gamma \frac{p}{\rho}  \equiv c_0^2 = \left ( \frac{\partial p}{\partial \rho} \right )_{\!\!s} .
\ee

\section{Ionization Contributions of all Elements in the Quiet Solar Atmosphere}
\label{sec-ion}

\begin{figure}[ht]
\centering
\includegraphics[scale=0.8]{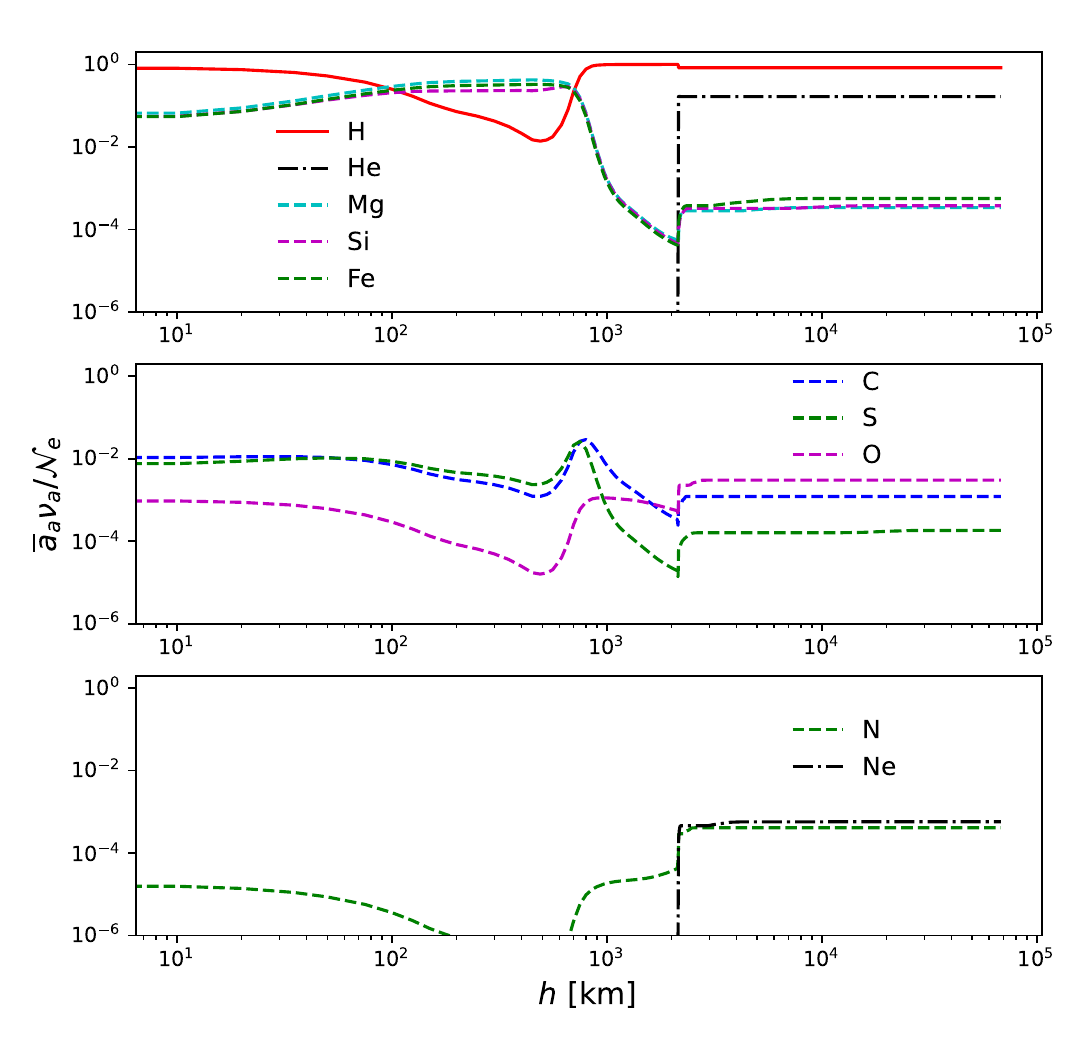}
\caption{Ionization contribution $\overline{a}_a\nu_a/\mathcal{N}_e$ see \Eqref{Electron_Density} of all 10 chemical elements in the solar plasma cocktail for the AL08 profile (same as \Fref{AL08} below).
Top: elements with dominant or significant contribution;
middle: elements with moderate contribution; 
below: elements with negligible contribution.
This type of sorting serves to illustrate which chemical elements may be left out of further research and on which should the focus turn on more closely.}
\label{ne-AL08}
\end{figure}

The contribution of each chemical element in the plasma cocktail to its ionization is next described more thoroughly.
In \Fref{AL08} below the ionization contribution of selected chemical elements is shown only due to clarity considerations.
Following the same axes and notions and height profile by \citet{Avrett:08} in \Fref{ne-AL08} the ionization contribution of all used in the calculation elements is shown in 3 separate vertically placed panels.
One can see that in chromosphere between 100~km and 1000~km
the contribution of Mg, Si and Fe dominates over the hydrogen.
Outside of this interval one can consider the solar chromosphere as a cocktail of partially ionized H and neutral He, i.e. $r_{0,\mathrm{He}}\approx 1$.
For this special compound for sound velocity and the polytropic index
the general equations are solved and we have the analytical result
\be
\tilde{\gamma}_{_\mathrm{H,\,He}}=c_0^2/(p/\rho),
\qquad
c_0^2\approx\frac{2c_p+(1-\alpha)\alpha(c_p+\iota)^2
+\overline a_{_\mathrm{He}}(2-\alpha)c_p}
{2c_v+(1-\alpha)\alpha\left[(c_v+\iota)^2+c_v\right]
+\overline a_{_\mathrm{He}}(2-\alpha)c_v}
\frac{p}{\rho},
\quad
\alpha\equiv r_{_{1,\, \mathrm{H}}},
\quad
I\equiv I_{1,\mathrm{H}},
\quad
\iota \equiv \frac{I}{T}.
\label{c0_Helium}
\ee
where $\alpha$ is the hydrogen degree of ionization with ionization potential $I$.
Let us mention that for partial ionization at low temperatures
when
$(1-\alpha)\alpha I\gg T$ polytropic index $\gamma\rightarrow 1$
as it was initially supposed by Newton.
In conclusion the deep minimum of the height dependence of the polytropic index depicted in Fig.~\ref{gamma-AL08}
is a consequence of hydrogen ionization in the upper chromosphere
where helium is neutral and contribution of metals in electron density is negligible.
The top panel shows the main contributing elements to the ionization, each one of these is either dominant or significant at a certain height from the Sun.
Next in the middle panel are the elements that have a little contribution, each several percent at most.
Finally below are the chemical elements with hardly noticeable contribution to the ionization.

These results show that N and Ne can be safely excluded from similar future studies.
Instead of having low contributing chemical elements, it is much more reasonable excitation levels of the remaining chemical elements to be included in the Saha equation \Eqref{Saha} with their energy levels added to the ionization potentials $I_{i,a}$.


\bibliography{zeta}

\begin{thebibliography}{}
\expandafter\ifx\csname natexlab\endcsname\relax\def\natexlab#1{#1}\fi
\providecommand{\url}[1]{\href{#1}{#1}}
\providecommand{\dodoi}[1]{doi:~\href{http://doi.org/#1}{\nolinkurl{#1}}}
\providecommand{\doeprint}[1]{\href{http://ascl.net/#1}{\nolinkurl{http://ascl.net/#1}}}
\providecommand{\doarXiv}[1]{\href{https://arxiv.org/abs/#1}{\nolinkurl{https://arxiv.org/abs/#1}}}

\bibitem[{Avrett \& Loeser(2008)}]{Avrett:08}
Avrett, E.~H., \& Loeser, R. 2008, ApJS, 175, 229, \dodoi{10.1086/523671}

\bibitem[{Basu \& Mandel(2004)}]{Basu:04}
Basu, S., \& Mandel, A. 2004, ApJ, 617, L155, \dodoi{10.1086/427435}

\bibitem[{{Baturin} {et~al.}(2022){Baturin}, {Oreshina}, {D{\"a}ppen},
  {Ayukov}, {Gorshkov}, {Gryaznov}, \& {Iosilevskiy}}]{Baturin:22}
{Baturin}, V.~A., {Oreshina}, A.~V., {D{\"a}ppen}, W., {et~al.} 2022, A\&A,
  660, A125, \dodoi{10.1051/0004-6361/202141873}

\bibitem[{{Brekke}(1993)}]{HRTS}
{Brekke}, P. 1993, ApJS, 87, 443, \dodoi{10.1086/191810}

\bibitem[{{Curdt, W.} {et~al.}(2001){Curdt, W.}, {Brekke, P.}, {Feldman, U.},
  {Wilhelm, K.}, {Dwivedi, B. N.}, {Sch\"uhle, U.}, \& {Lemaire, P.}}]{SUMER}
{Curdt, W.}, {Brekke, P.}, {Feldman, U.}, {et~al.} 2001, A\&A, 375, 591,
  \dodoi{10.1051/0004-6361:20010364}

\bibitem[{{Dayeh} \& {Livadiotis}(2022)}]{Dayeh:22}
{Dayeh}, M.~A., \& {Livadiotis}, G. 2022, ApJL, 941, L26,
  \dodoi{10.3847/2041-8213/aca673}

\bibitem[{{Donkov} {et~al.}(2021){Donkov}, {Stefanov}, {Veltchev}, \&
  {Klessen}}]{Donkov:21}
{Donkov}, S., {Stefanov}, I.~Z., {Veltchev}, T.~V., \& {Klessen}, R.~S. 2021,
  MNRAS, 505, 3655, \dodoi{10.1093/mnras/stab1572}

\bibitem[{Doorsselaere {et~al.}(2011)Doorsselaere, Wardle, Zanna, Jansari,
  Verwichte, \& Nakariakov}]{Van_Doorsselaere:11}
Doorsselaere, T.~V., Wardle, N., Zanna, G.~D., {et~al.} 2011, ApJL, 727, L32,
  \dodoi{10.1088/2041-8205/727/2/l32}

\bibitem[{{Houston} {et~al.}(2018){Houston}, {Jess}, {Asensio Ramos}, {Grant},
  {Beck}, {Norton}, \& {Krishna Prasad}}]{Houston:18}
{Houston}, S.~J., {Jess}, D.~B., {Asensio Ramos}, A., {et~al.} 2018, ApJ, 860,
  28, \dodoi{10.3847/1538-4357/aab366}

\bibitem[{Jacobs \& Poedts(2011)}]{Poedts:11}
Jacobs, C., \& Poedts, S. 2011, AdSpR, 48, 1958 ,
  \dodoi{10.1016/j.asr.2011.08.015}

\bibitem[{Kramida {et~al.}(2022)Kramida, {Yu.~Ralchenko}, Reader, \& {and NIST
  ASD Team}}]{NIST_ASD}
Kramida, A., {Yu.~Ralchenko}, Reader, J., \& {and NIST ASD Team}. 2022, {NIST
  Atomic Spectra Database (ver. 5.10), [Online]. Available:
  {\tt{https://physics.nist.gov/asd}} [2023, September 4]. National Institute
  of Standards and Technology, Gaithersburg, MD.},
  \dodoi{doi.org/10.18434/T4W30F}

\bibitem[{Landau \& Lifshitz(1980)}]{LL5}
Landau, L.~D., \& Lifshitz, E.~M. 1980, Course of Theoretical Physics, Vol.~5,
  Statistical Physics. Part 1 (Oxford: Butterworth-Heineman)

\bibitem[{{Livadiotis} \& {McComas}(2023)}]{Livadiotis:23}
{Livadiotis}, G., \& {McComas}, D.~J. 2023, ApJ, 954, 72,
  \dodoi{10.3847/1538-4357/ace1e3}

\bibitem[{Mishonov {et~al.}(2021a)Mishonov, Dimitrova, \& Varonov}]{PhysA}
Mishonov, T.~M., Dimitrova, I.~M., \& Varonov, A.~M. 2021a, PhyA, 563, 125442,
  \dodoi{https://doi.org/10.1016/j.physa.2020.125442}

\bibitem[{{Mishonov} {et~al.}(2021b){Mishonov}, {Dimitrova}, \&
  {Varonov}}]{ApJ:21}
{Mishonov}, T.~M., {Dimitrova}, I.~M., \& {Varonov}, A.~M. 2021b, ApJ, 916, 18,
  \dodoi{10.3847/1538-4357/ac0629}

\bibitem[{Mishonov \& Varonov(2020)}]{Pade}
Mishonov, T.~M., \& Varonov, A.~M. 2020, ApNM, 157, 291,
  \dodoi{https://doi.org/10.1016/j.apnum.2020.06.007}

\bibitem[{Nicolaou {et~al.}(2023)Nicolaou, Livadiotis, \&
  McComas}]{Nicolaou:23}
Nicolaou, G., Livadiotis, G., \& McComas, D.~J. 2023, ApJ, 948, 22,
  \dodoi{10.3847/1538-4357/acbf33}

\bibitem[{Nicolaou {et~al.}(2020)Nicolaou, Livadiotis, Wicks, Verscharen, \&
  Maruca}]{Nicolaou:20}
Nicolaou, G., Livadiotis, G., Wicks, R.~T., Verscharen, D., \& Maruca, B.~A.
  2020, ApJ, 901, 26, \dodoi{10.3847/1538-4357/abaaae}

\bibitem[{Pang {et~al.}(2022)Pang, Geng, Wang, Cao, Deng, Duan, Li, Jia, \&
  Xu}]{Pang:22}
Pang, X., Geng, X., Wang, S., {et~al.} 2022, ApJ, 940, 120,
  \dodoi{10.3847/1538-4357/ac9d2d}

\bibitem[{Petrie {et~al.}(2007)Petrie, Blokland, \& Keppens}]{Petrie:07}
Petrie, G. J.~D., Blokland, J. W.~S., \& Keppens, R. 2007, ApJ, 665, 830,
  \dodoi{10.1086/519276}

\bibitem[{Prasad {et~al.}(2018)Prasad, Raes, Doorsselaere, Magyar, \&
  Jess}]{Prasad:18}
Prasad, S.~K., Raes, J.~O., Doorsselaere, T.~V., Magyar, N., \& Jess, D.~B.
  2018, ApJ, 868, 149, \dodoi{10.3847/1538-4357/aae9f5}

\bibitem[{Saha(1921)}]{Saha:21}
Saha, M.~N. 1921, Proc. R. Soc. Lond. A, 99, 135,
  \dodoi{10.1098/rspa.1921.0029}

\bibitem[{Totten {et~al.}(1995)Totten, Freeman, \& Arya}]{Totten:95}
Totten, T.~L., Freeman, J.~W., \& Arya, S. 1995, JGRA, 100, 13,
  \dodoi{https://doi.org/10.1029/94JA02420}

\bibitem[{{Vashalomidze} {et~al.}(2019){Vashalomidze}, {Zaqarashvili}, \&
  {Kukhianidze}}]{Vashalomidze:19}
{Vashalomidze}, Z.~M., {Zaqarashvili}, T.~V., \& {Kukhianidze}, V.~D. 2019, Ap,
  62, 69, \dodoi{10.1007/s10511-019-09565-8}

\bibitem[{{Vernazza} {et~al.}(1981){Vernazza}, {Avrett}, \& {Loeser}}]{VAL:81}
{Vernazza}, J.~E., {Avrett}, E.~H., \& {Loeser}, R. 1981, ApJS, 45, 635,
  \dodoi{10.1086/190731}

\bibitem[{{Wang} {et~al.}(2015){Wang}, {Ofman}, {Sun}, {Provornikova}, \&
  {Davila}}]{Wang:15}
{Wang}, T., {Ofman}, L., {Sun}, X., {Provornikova}, E., \& {Davila}, J.~M.
  2015, ApJL, 811, L13, \dodoi{10.1088/2041-8205/811/1/L13}

\bibitem[{Wynn(1966)}]{Wynn:66}
Wynn, P. 1966, NuMat, 8, 264.
\newblock \url{https://api.semanticscholar.org/CorpusID:123789548}

\end{thebibliography}

\end{document}